\documentclass[pra,aps,twocolumn,showpacs]{revtex4}
\usepackage{times,epsfig,amssymb,amsfonts,amsmath}

\newcommand{\ket}[1]{|#1\rangle}
\newcommand{\bra}[1]{\langle #1|}

\begin{document}

\title{Multipartite quantum correlation, spatially anisotropic coupling, and finite temperature effects in a triangular Ising system with tunable interactions}

\author{Jun Ren$^1$}
\author{Fang-Man Liu$^1$}
\author{Yan-Chao Li$^2$}
\author{Li-Hang Ren $^1$}
\author{Z. D. Wang$^3$}
\email{zwang@hku.hk}
\author{Yan-Kui Bai$^{1,3}$}
\email{ykbai@semi.ac.cn}

\affiliation{$^1$ College of Physics and Hebei Key Laboratory of Photophysics Research and Application, Hebei Normal University, Shijiazhuang, Hebei 050024, People's Republic of China\\
	$^2$ College of Materials Science and Opto-Electronic Technology, University of Chinese Academy of Sciences, Beijing 100049, People's Republic of China\\
$^3$ Guangdong-Hong Kong Joint Laboratory of Quantum Matter, Department of Physics, and HK Institute of Quantum Science \& Technology, The University of Hong Kong, Pokfulam Road, Hong Kong, China}

\begin{abstract}
We investigate multipartite quantum correlation (MQC), spatially anisotropic coupling, and finite temperature effects in a triangular Ising system with tunable interactions using the exact diagonalization method. We demonstrate that spatially anisotropic coupling serves as an effective means to modulate MQC in the antiferromagnetic ground state, which is achievable with current experimental technologies. Moreover, we explore the interplay between MQC and spatially anisotropic coupling in the Ising system at finite temperatures. Our findings reveal a three-way trade-off relationship among high MQC, robust thermal stability, and anisotropic strength in the triangular Ising system with antiferromagnetic interactions, though the MQC in the ferromagnetic case is quite susceptible to temperature changes. These insights contribute to our understanding of ground state properties and MQC modulation in quantum many-body systems.
\end{abstract}


\maketitle

\section{Introduction}
Quantum correlation in the ground state of many-body systems is a kind of important physical resource for quantum information processing \cite{amico08rmp,horod09rmp,modi12rmp,bera18rpp}. Frustrated spin systems, in particular, have garnered significant attention due to their highly degenerate ground states, which can result in multipartite quantum correlations (MQCs) \cite{sach99book,chiara18rpp}.
The most basic form of spin frustration can be observed in a triangular lattice with three spins and antiferromagnetic interactions \cite{binder86rmp}, where the spins cannot organize into the preferred antiparallel pattern, resulting in a degenerate manifold of ground states due to the competition between interactions and lattice geometry.
The ground state of many-body spin systems can be in different phases at zero temperature \cite{sach99book}, and the MQC serves as an effective tool for characterizing the properties of ground state in different quantum phases \cite{weitc05pra,olivei06pra,olivei06prl,facchi10njp,mon10pra,giampa13pra,sunzy14pra,hofm14prb,Bayat17prl,Yamasaki18pra,haldar20prb}. Recent theoretical and experimental research has demonstrated that spin lattices with spatially anisotropic triangular configurations possess unique phases and potential applications in quantum computing \cite{Yoshida15np,Zhang16prl,Ito16prb,Keles18prl,Tala20prapp}.

Quantitative characterization of MQC is a fundamental problem in quantum information processing. Entanglement monogamy in multipartite systems implies that quantum entanglement cannot be freely shared among different subsystems \cite{bennett96pra} and the quantitative monogamous relations can be used to construct the measures or indicators for genuine MQC \cite{coffman00pra,osborne06prl,christandl04jmp,fan07pra,byw07pra,kimjs09pra,bai06pra,byw09pra,kimjs10jpa,cornelio10pra,bxw14prl,songbai16pra,ren21npj,bai22pra}. Using the tripartite quantum correlation based on the monogamy relation of negativity \cite{fan07pra,vidal02pra}, Rama Koteswara Rao \emph{et al} explored the ground state of quantum transverse Ising model in a fixed triangular configuration through a nuclear magnetic resonance (NMR) system \cite{rkrao13pra}. They found that, unlike the bipartite quantum correlation, multipartite quantum correlation can differentiate between frustrated and non-frustrated regimes in the ground state of a three-spin system, with higher MQC values in the non-frustrated regime compared to the frustrated one \cite{rkrao13pra}. In addition, it is worth noting that a system's ground state property is highly dependent on its anisotropy, such as anisotropic Heisenberg chains \cite{wang01pla,kamta02prl,zhou03pra,zhang05pra} and spatially anisotropic triangular lattices \cite{yunoki06prb,eckardt10epl,hauke13prb}. Consequently, it is of interest to study the ground state properties in triangular Ising systems with tunable spatially anisotropic couplings and investigate a more general relationship between MQC and anisotropy in both ferromagnetic and antiferromagnetic ground states of Ising systems, particularly in systems beyond the three-spin case.

On the other hand, it is also essential to further analyze the MQC in tunable Ising systems at finite temperatures, as MQC plays a vital role in quantum information processing \cite{cory98prl,knill98prl,rauss01prl,datta08prl,lanyon08prl,ren21arxiv,monz09prl,lanyon09natp}. Over the past two decades, significant efforts have been made to study thermal quantum entanglement or quantum correlation in spin systems, as highlighted in the review papers \cite{amico08rmp,bera18rpp}. In this context, we aim to investigate the intrinsic relationship between MQC and spatially anisotropic strength in tunable Ising systems at finite temperatures and explore an effective and feasible method for MQC modulation using current experimental technologies \cite{kim10nat,Struck11sci,wang18natnano,niskanen07sci,Ryazanov01prb,Grosz11prb,Kosior13pra,Seo13prl}. It remains an open question whether there exists an inherent connection between high MQC, strong thermal robustness, and spatially anisotropic strength, even in the simplest three-spin case.

In this paper, we investigate the multipartite quantum correlations (MQC), spatially anisotropic coupling, and finite temperature effects in a triangular Ising system with tunable interactions up to 15 spins, using the exact diagonalization method. The paper is organized as follows: In Section II, we first introduce the relevant background knowledge, including the tunable triangular Ising system and the MQC measure. We then examine the properties of MQC as the spatial anisotropy changes in both ferromagnetic and antiferromagnetic ground states. In Section III, we analyze the effects of finite temperature on MQC and explore the inherent relationship among the three key elements, namely, high MQC, its thermal robustness, and the anisotropic strength in the antiferromagnetic case. We also explain why MQC in the ferromagnetic case is highly susceptible to temperature. Finally, in Section IV, we discuss the experimental possibilities for MQC modulation via spatially anisotropic coupling in a system of cold atoms trapped in an optical lattice and provide a brief conclusion.

\section{The MQC property in the ground state of a tunable triangular Ising system}

\subsection{The tunable triangular Ising system and the MQC characterization}

In this work, we address the two-dimensional triangular Ising lattices as shown in Fig.~1, for which the Hamiltonian with only nearest-neighbor couplings has the form
\begin{equation}\label{1}
H=\sum_{\left\langle i,j\right\rangle } J_{ij}S_i^x S_j^x+h \sum_{i} S_i^z,
\end{equation}
where $J_{ij}$ is the coupling strength between the $i$th and the $j$th spins, $h$ means the strength of the transverse field, $S_i^x$ and $S_i^z$ are spin-1/2 operators with the subscript being the spin located at the vertex  $i$ in the two-dimensional triangular lattices. As shown in the figure, the nearest-neighbor coupling $J_{ij}$s in different directions are indicated by blue, red and green line segments,  for which their magnitudes can be expressed as $\omega J$, $\eta J$ and $J$, respectively, with $\omega$ and $\eta$ being two nonnegative tunable parameters. It is noted that the triangular structure with fixed couplings in the NMR system \cite{rkrao13pra} is covered by the Hamiltonian. When $J>0$, the system possesses antiferromagnetic interactions and may result in frustration, while the interactions are ferromagnetic in the case of $J<0$ and the system has no frustration. The eigenstates and eigenenergies for the Ising system can be solved by an exact diagonalization approach. Hereafter we set the parameter $h=1$ for simplicity.

\begin{figure}
	\epsfig{figure=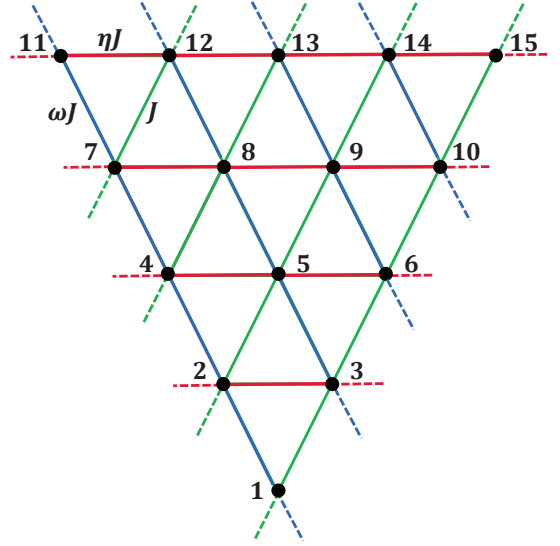,width=0.4\textwidth}
	\caption{(Color online) The schematic diagram of two-dimensional triangular Ising lattices with the tunable couplings (blue lines and red lines) via spatially anisotropic parameters $\omega$ and $\eta$.}
\end{figure}

Entanglement negativity \cite{vidal02pra,plenio05prl,zyc98pra} is an important entanglement measure, which is computable for bipartite mixed states and can be defined as \cite{vidal02pra}
\begin{equation}\label{2}
N(\rho_{AB})=\|\rho_{AB}^{T_A}\|_1-1,
\end{equation}
where $\|\cdot\|_1$ denotes the trace norm and equals the sum of the moduli of eigenvalues for the partial transposed matrix $\rho_{AB}^{T_A}$. Entanglement monogamy can be utilized to constitute the measure of multipartite quantum correlation \cite{horod09rmp}. Ou and Fan proved the monogamy property of squared negativities in pure multiqubit states, and show genuine tripartite quantum correlation in a three-qubit pure state $\ket{\psi}_{ABC}$ can be characterized by the residual correlation $\pi_3^{A|BC}= N_{A|BC}^2-N_{AB}^2-N_{AC}^2$ \cite{fan07pra} in which $N_{A|BC}$ is the bipartite negativity in the partition $A|BC$ and $N_{Aj}$ quantifies the two-qubit entanglement with $j=\{B, C\}$.

For the ground state of the tunable triangular Ising system with $N$ spins, the MQC can be characterized by the residual entanglement of the squared negativities. Here, we make use of a modified multipartite quantum correlation measure in an $N$-qubit state $\ket{\Psi}$ via the monogamous property of squared negativities
\begin{equation}
	T_{N}(A_i)=[N_{A_i|\bar{A}_i}^{2}-\sum_{j\neq i}N_{A_i|A_j}^{2}]^{\frac{1}{2}},
\end{equation}
where $N_{A_i|\bar{A}_i}$ is the bipartite negativity between qubit $A_i$ and the other subsystems $\bar{A}_i$, $N_{A_i|A_j}$ is the two-qubit negativity between the $i$th and $j$th qubits, and the letter $A_i$ in the parenthesis for the MQC $T_N(A_i)$ means that the qubit $A_i$ is the central qubit in the $N$-partite correlation.

\begin{figure*}
	\epsfig{figure=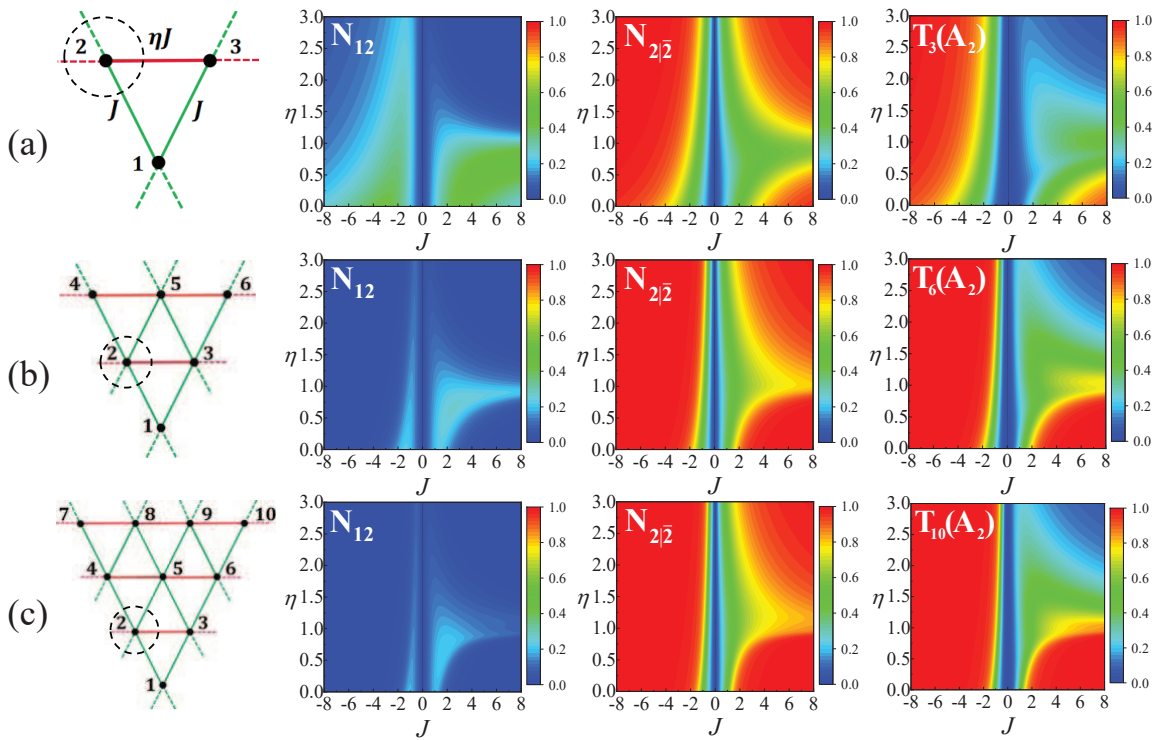,width=0.85\textwidth}
	\caption{(Color online) The schematic diagrams of the tunable triangular Ising lattices, two-spin quantum correlation $N_{12}$, bipartite multi-spin correlation $N_{2|\bar{2}}$, and the MQC $T_N$ as the functions of the spatially anisotropic parameter $\eta$ and interaction strength $J$ in the ground state. (a) Three-spin case: from left to right, the four panels are schematic diagram for three-spin lattice with one tunable interaction, two-spin quantum correlation $N_{12}$, bipartite three-spin correlation $N_{2|\bar{2}}$, the MQC $T_3(A_2)$, respectively. (b) Six-spin case: the schematic diagram for the tunable lattices with six spins, and three correlation measures $N_{12}$, $N_{2|\bar{2}}$ and $T_6(A_2)$. (c) Ten-spin case: schematic diagram for the ten-spin lattices, and three quantum correlations $N_{12}$,$N_{2|\bar{2}}$ and $T_{10}(A_2)$, respectively.}
\end{figure*}

\subsection{The MQC property and spatial anisotropy in the ground states of tunable triangular Ising system}

We first consider the triangular Ising model with one tunable coupling parameter $\eta$ by setting the other parameter $\omega=1$, which means that the nearest-neighbor couplings with the magnitude $J_{ij}=\eta J$ in horizontal direction is tunable, as shown in Fig. 2. In experiments, the tunable triangular configuration with two equally coupled legs has been realized in ion traps \cite{kim10nat} and optical lattices \cite{Struck11sci}.

In Fig. 2(a), we first analyze the three-spin case, where the first panel is the schematic diagram with the tunable coupling $\eta J$ (red line) and the central qubit with a dotted circle is the spin number $2$. The ground state belongs to the class of tetrahedral states \cite{Carteret00jpa,Brun01pla} and can be used to perform the controlled teleportation \cite{Li14pra}. In the second panel of Fig. 2(a), the negativity $N_{12}$ is plotted as a function of the tunable parameter $\eta$ and the interaction strength $J$. As shown in the panel, the value of $N_{12}$ is small, where $N_{12}$ decreases along with the increase of $\eta$ for a given $J$ in the ferromagnetic regime and the negativity is almost zero when $\eta>1$ in antiferromagnetic regime. The third panel of Fig. 2(a) shows the bipartite negativity $N_{2|\bar{2}}$ as the function of $\eta$ and $J$, which quantifies the bipartite quantum correlation between qubit $2$ and other qubits $\bar{2}$. As shown in the panel, $N_{2|\bar{2}}$ in ferromagnetic regime has the higher values and is insensitive to the anisotropic parameter $\eta$ for larger $J$, while $N_{2|\bar{2}}$ in antiferromagnetic regime is sensitive to the anisotropy and has lower values for the configuration near the isotropic lattices. In the last panel of Fig. 2(a), the tripartite quantum correlation $T_{3}(A_2)=[ N_{2|13}^{2}-N_{12}^{2}-N_{23}^{2}] ^{1/2}$ is plotted, where the MQC in the ferromagnetic ground state (away from the critical point $J=0$) has a relatively large value and is insensitive to the spatially anisotropic parameter $\eta$, but the correlation in the antiferromagnetic case is quite different and decreases along with the parameter $\eta$ increases. By comparing $N_{2|\bar{2}}$ with $T_{3}$, we obtain that the bipartite correlation $N_{2|\bar{2}}$ in ferromagnetic regime is dominated by MQC $T_3$ and the case of antiferromagnetic regime with small $\eta$ is similar, while $N_{2|\bar{2}}$ in antiferromagnetic regime with large $\eta$ is mainly composed of two qubit negativities. According to Fig. 2(a), the MQC $T_{3}$ can be effectively modulated from zero to almost the maximum via tuning the anisotropy parameter $\eta$.

\begin{figure*}
	\epsfig{figure=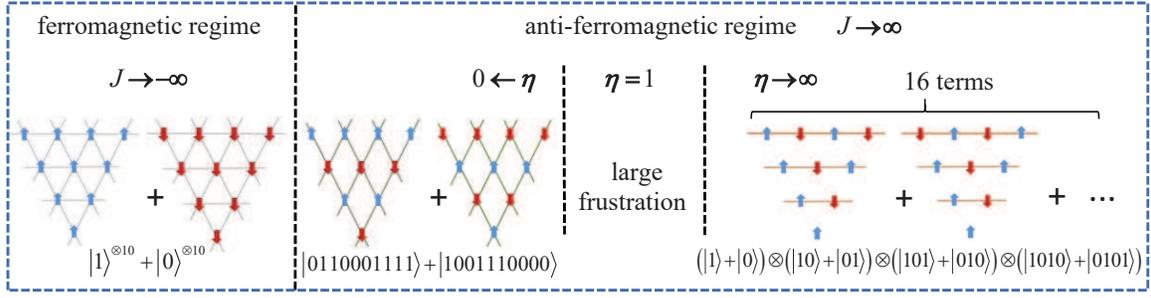,width=0.85\textwidth}
	\caption{(Color online) The schematic diagram for the relationship between the spatially anisotropic coupling and quantum correlation structure in the ground state of ten spins far away from the critical point, where the left part is the ferromagnetic ground state being insensitive to the anisotropic parameter $\eta$ and the right part is the antiferromagnetic ground state being sensitive to the anisotropy.}
\end{figure*}

In Figs. 2(b) and 2(c), we further analyze correlation properties in the tunable triangular Ising lattices with six and ten spins, respectively, where the couplings (red lines) in horizontal direction are tunable via the spatially anisotropic parameter $\eta$ and the central qubit is still the spin number 2 as shown in the first panels of the two figures. In the second panel of Fig. 2(b), the two-spin entanglement $N_{12}$ is plotted as a function of the parameters $\eta$ and $J$, where both ferromagnetic and antiferromagnetic regimes have small values of two-qubit quantum correlation. In the third panel of Fig. 2(b), we plot the bipartite entanglement $N_{2|\bar{2}}$ for the lattices of six spins, which is similar to the three-spin case and $N_{2|\bar{2}}$ in the ferromagnetic regime is insensitive to $\eta$ for larger $J$ and the bipartite multi-spin correlation has a lower value for the configuration near isotropic lattices. The MQC $T_6(A_2)$ in the ground state of the triangular lattices is plotted in the last panel of Fig. 2(b), where the anisotropic parameter $\eta$ has little influence on the MQC in the ferromagnetic ground state (away from the critical point) but can modulate the MQC in the antiferromagnetic region. The case for the lattices with ten spins was shown in Fig. 2(c), where the correlations $N_{12}$, $N_{2|\bar{2}}$, and $T_{10}(A_2)$ have the similar behaviors along with the changes of the parameters $\eta$ and $J$ as those of six-spin case and the spatially anisotropic parameters $\eta$ can effectively modulate the MQC $T_{10}$ in antiferromagnetic ground state. In addition, along with the increase of spin numbers, the high quantum correlation areas of $N_{2|\bar{2}}$ and $T_N$ are enlarged.

The influence of the spatially anisotropic parameter $\eta$ on MQC $T_{N}$ is quite different in the ferromagnetic and antiferromagnetic ground states as shown in Fig. 2. Here, we analyze the reason and consider the regions of interaction strength $J$ far away from the critical point. As an example, we study the triangular lattices with ten spins. In the ferromagnetic regime, regardless of the anisotropic strength of $\eta$, the ground state of the system always tends to the state of all the spins being spin-up or spin-down at the same time. In the case of $J\rightarrow -\infty$, the ground state is just the ferromagnetic state (GHZ-type) $\ket{\Psi}=(\ket{0}^{\otimes {10}}+\ket{1}^{\otimes {10}})/\sqrt{2}$ with the maximally multipartite quantum correlation as shown in the left part of Fig.~3. Therefore, the anisotropic parameter $\eta$ does not change the property of the ferromagnetic ground state and has little influence on the MQC $T_{10}$  when the interaction strength $J$ in the nonfrustration regime has a larger value as exhibited in the last panel of Fig. 2(c). However, the MQC property of $T_{10}$ in the antiferromagnetic ground state is sensitive to the spatially anisotropic parameter $\eta$ as shown in the right part of Fig. 3. When the parameter $\eta\rightarrow 0$ and the coupling strength $J\rightarrow \infty$, the interactions in horizontal direction disappear and there is no frustration in the lattices. As shown in the figure, the ground state is a kind of GHZ state $\ket{\Phi}=(\ket{011000111}+\ket{100111000})/\sqrt{2}$, for which the MQC has the maximum. When $\eta=1$ with a larger $J$, the antiferromagnetic ground state has an isotropic structure and tends to the equal probability superposition of a lot of degenerate states. In this case, the ground state has the large frustration since one-third nearest-neighbor spin pairs cannot order in the favored antiparallel pattern and its MQC is nonzero but not maximal. Along with the increases of the parameter $\eta$, the anisotropy will affect the competition among these degenerate states, and thus change the MQC in the ground state. In the case of $\eta\rightarrow \infty$, the antiferromagnetic ground state becomes a tensor product state, which can be written as $\ket{\Psi'}=(\ket{1}+\ket{0})\otimes(\ket{10}+\ket{01})\otimes(\ket{101}+\ket{010})\otimes(\ket{1010}+\ket{0101})/4$, which results in the MQC $T_{10}(A_2)$ being zero. It should be noted that our analysis on the relation between the anisotropy of Ising lattices and the MQC in the ground state is not limited to the case of ten spins, and the qualitative results still hold for the case of more spins.

According to the numerical results in Fig. 2 and our qualitative analysis shown in Fig. 3, we can obtain that, regardless of the number of spins, the anisotropy parameter $\eta$ has little influence on the MQC $T_{N}$ in the ground state with ferromagnetic interaction away from the critical point, but for the antiferromagnetic case ($J>0$), the anisotropic parameter $\eta$ can greatly change the ground state property and its MQC $T_N$. In particular, when $\eta$ tends to zero with $J\rightarrow \infty$, the ground state tends to the GHZ state which has the maximal MQC. When $\eta\rightarrow \infty$ with large $J$, the ground state tends to a tensor product state and its MQC is zero. In the case of $\eta$ close to 1, the ground state has large frustration for which the MQC has a medium value. Moreover, although the modulation effect of the MQC has a little difference along with the increase of spin numbers, the MQC in the antiferromagnetic ground state (with a proper value of $J$) can be continuously adjusted on a scale near $(0,1)$.

In order to illustrate our results, we further plot the MQC $T_N(A_2)$ as a function of parameter $J$ in the ground state of Ising lattices with the anisotropic parameter $\eta=0.2, 1$, and $2.5$, where the numbers of spins are chosen to be $3,6,10$, and $15$ respectively. As shown in Fig. 4, the anisotropy has little influence on the high values of the MQCs in the ferromagnetic regime ($J<0$), while the values of the MQCs in the antiferromagnetic regime ($J>0$) decrease along with the increase of $\eta$. For example, the value of the MQC $T_{15}$ with $J=6$ is modulated from almost the maximum $1$ to the lower value $0.15$ when the observer changes the anisotropic parameter $\eta$ from $0.2$ to $2.5$. In addition, the MQC $T_{N}(A_2)$ in both two regimes increases along with the increase of the number of lattice spins.

\begin{figure*}
	\epsfig{figure=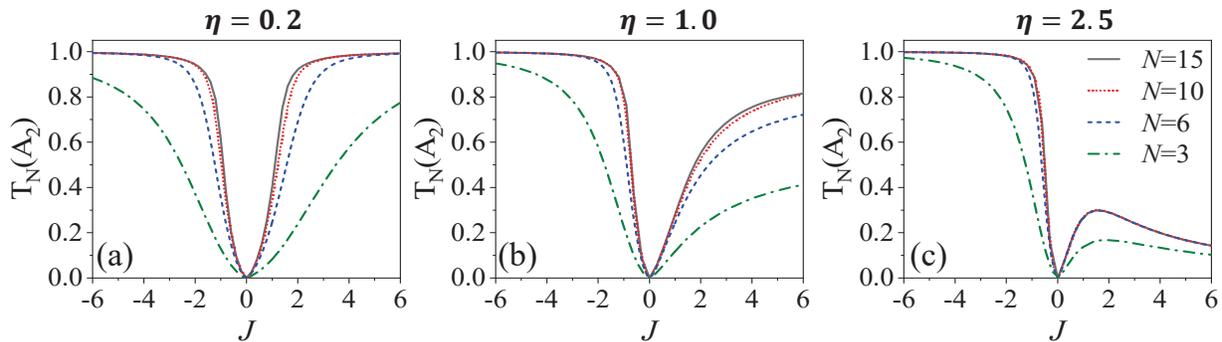,width=0.9\textwidth}
	\caption{(Color online) The MQC $T_N(A_2)$ as a function of parameter $J$ in the ground state of Ising lattices with the anisotropic parameter $\eta=0.2, 1$, and $2.5$, where the numbers of spins are chosen to be $3,6,10$, and $15$ respectively.}
\end{figure*}

We have studied the relation between the MQC property and spatial anisotropy in the triangular Ising lattices with one asymmetric interaction. In practical quantum systems, there also exist completely spatially anisotropic coupling systems, such as the dipole-dipole interaction Heisenberg model \cite{Keles18prl}, the tunable artificial spin ice structure \cite{Tala20prapp}, and the superconducting qubit architecture \cite{Grosz11prb}, and so on. In the next section, we will analyze the completely spatially anisotropic triangular Ising system and explore the thermal robustness of the MQC.

\section{The MQC and its thermal robustness in the spatially anisotropic Ising system}

Characterization of thermal quantum resources is an important issue in quantum information processing, and considerable efforts have been made to study the thermal bipartite entanglements or quantum correlations in spin systems where the effects of magnetic field and temperature are analyzed, see, for example, Refs. \cite{wang01pla,kamta02prl,zhou03pra,zhang05pra,arnesen01prl,wang01pra,wang02pra,sun05njp,abliz06pra,ww06pra,hide09prl,park17prl,Shim08prb} and the review papers \cite{amico08rmp,bera18rpp}. Since multipartite quantum correlation is a kind of important physical resource \cite{vedral04njp,Nakata09pra,Campbell13njp,sun19pra}, it is necessary to investigate the property of the MQC at finite temperatures. Here, we will focus on the inherent relation among the MQC, its thermal robustness and spatial anisotropy in the tunable Ising system with the triangular configuration.

The thermal state under the equilibrium at a finite temperature $T$ can be written as \cite{sach99book} $\rho(T)=\frac{1}{Z} \mbox{exp}(-\beta H)$ in which $Z=\mbox{tr}[\mbox{exp}(-\beta H)]$ is the partition function with $H$ being the system Hamiltonian and  $\beta=1/k_{B}T$ (for simplicity, the Boltzmann constant is set to be $k_B=1$). In the basis of eigenvectors, the thermal state of the tunable Ising system can be further expressed as
\begin{equation}
\rho(T)=\frac{1}{Z} \sum_{i}e^{-E_i/T}\ket{e_{i}}\bra{e_{i}},
\end{equation}
where $E_i$s are energy eigenvalues of the Hamiltonian given in Eq. (1) and the coefficient $e^{-E_i/T}$ determines the proportion of the eigenstate $\ket{e_i}$ in the mixed state.

\begin{figure*}
	\epsfig{figure=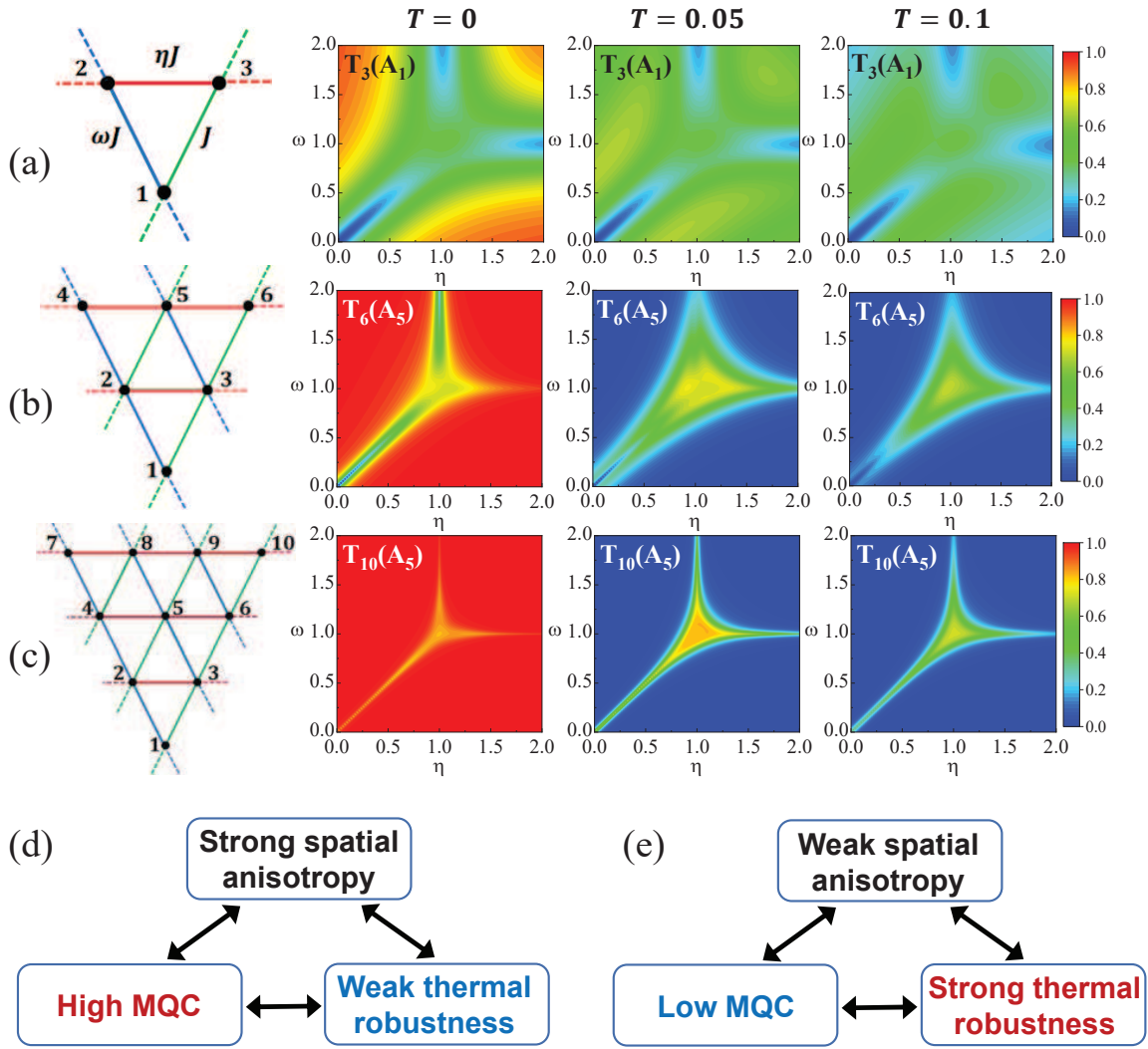,width=0.85\textwidth}
	\caption{(Color online) The MQC and its thermal robustness in the spatially anisotropic triangular lattices. (a)-(c) The anisotropic lattices with spin numbers $N=3, 6, 10$ and the MQC $T_N$ as the function of anisotropic parameters $\omega$ and $\eta$ with the coupling strength $J=6$ at the temperatures $T=0, 0.05$, and $0.1$, respectively. (d), (e) The three-way trade-off relations among the MQC, its thermal robustness and the anisotropic strength in the triangular Ising system with antiferromagnetic interactions.}
\end{figure*}

The triangular Ising lattices which we consider in this section are completely spatially anisotropic, and the nearest-neighbor couplings are $J_{ij}=\omega J$, $\eta J$ and $J$, respectively, in which the asymmetric couplings are tunable via the parameters $\omega$ and $\eta$. In order to analyze the finite temperature effect on the MQC, we choose the coupling strength $J=6$ away from the critical point and consider three typical temperatures $T=0, 0.05$, and $0.1$, respectively. In Fig. 5, we plot the MQC $T_N$ as a function of the anisotropic parameters $\omega$ and $\eta$  in the antiferromagnetic regime at different temperatures, where the spin numbers of the Ising lattices are $3$, $6$ and $10$ with the schematic diagrams shown in the left panels of Figs. 5(a)-(c). For the three-spin case, the MQC $T_3(A_1)$ at zero temperature is plotted as a function of the parameters $\omega$ and $\eta$ in the second panel of Fig. 5(a), where the lattice structure with larger anisotropy (orange and red regions) has the high value of the MQC, the structure with two smaller equal couplings (blue regions) has the low value of $T_3$, and the structure close to the isotropy (green central region) has the medium value. Along with the number of spins increases ($N=6$ and $N=10$), the area of high MQC (red regions) enlarges for the strong spatially anisotropic lattices, the low MQC (green and yellow regions) still corresponds to the structures with two smaller equal couplings, and the MQC for the structure close to isotropic lattices (central regions) increases as shown in the second panels in Figs. 5(b) and 5(c). At the finite temperature $T=0.05$, the MQC $T_N$ in the structure with larger anisotropy decreases rapidly but the MQC decreases slowly in the structure with weak spatial anisotropy (two smaller equal couplings and approximate isotropic couplings) as shown in the third panels of Figs. 5(a)-5(c). Moreover, along with the spin number increases, the MQC $T_N$ is more sensitive to the anisotropy and the area of strong thermal robustness is reduced. The case for the temperature $T=0.1$ is similar, where the MQC for the Ising lattices with weak spatial anisotropy is thermally robust and the area of nonzero MQC further decreases along the temperature as shown in the fourth panels of Figs. 5(a)-5(c). However, the thermal robustness of two-spin quantum correlations is quite different, and, in Appendix A, we analyze $\sum_i N_{5i}$ in the lattices of ten spins at finite temperature for which the configurations near the isotropic case have zero correlation.

According to the above analysis, we can obtain that there exists a three-way trade-off relation among high MQC, strong thermal robustness and the anisotropic strength in the triangular Ising system with antiferromagnetic interactions. The stronger the anisotropy of interactions, the higher the value of the MQC $T_N$, but the weaker the thermal robustness of the MQC. Conversely, the weaker the anisotropy of interactions, the lower the value of the MQC $T_N$, but the stronger the thermal robustness of the MQC. In Figs. 5(d) and 5(e), the schematic diagrams for three-way trade-off relations among the anisotropy, the MQC and its thermal robustness are plotted. Therefore, the anisotropy of the tunable Ising lattices is an important quantity to modulate the MQC and its thermal robustness. In particular, when both the relatively high MQC and strong thermal robustness are required in a certain task of quantum information processing, one can tune the spatially anisotropic interactions of triangular Ising system to a proper configuration.

Unlike the case of antiferromagnetic regime, the spatial anisotropy has little influence on the high MQC $T_N$ in the ground state of ferromagnetic case, since the ground state tends to the GHZ state as shown in the previous analysis of Fig. 3. However, the MQC $T_N$ in the ferromagnetic regime is very fragile at finite temperature \cite{ren21arxiv}. This property comes from the structure of thermal state in Eq. (5), which is closely related to the system eigenvalues and eigenstates. The mixed probability of component $\ket{e_i}$ is proportional to the value $e^{-E_i}$, and thus the eigenvector with the negative energy eigenvalue is dominant, while the one with positive eigenvalue has a small weight in the thermal state for a given lower temperature. As an example, we consider the case of three spins and plot the eigenvalue $E_i$ as a function of the coupling strength $J$ for the isotropic coupling $\eta=\omega=1$ in Fig. 6(a) and the anisotropic coupling $\eta=0.5$ and $\omega=1.5$ in Fig. 6(b). As shown in the figures, when $J<0$ in the ferromagnetic regime, the energy eigenvalues $E_0$ (purple solid line) and $E_1$ (green solid line) are negative and other eigenvalues are positive (black dotted lines), which leads to the thermal state being mainly comprised of the eigenvectors $\ket{e_0}$ and $\ket{e_1}$. Along with the increase of temperature, the ferromagnetic thermal state tends to be the mixture of the two eigenvectors with the equal probabilities due to the small gap between the ground state and the first excited state [see the inset of Fig.~6(a)], which results in a very low MQC and exhibit the weaker thermal robustness. In the isotropic case of antiferromagnetic thermal state ($J>0$), the ground state $\ket{e_0}$ is dominant in a certain temperature range and the MQC has the stronger thermal robustness, because the eigenenergies have a relatively large gap between $E_0$ and $E_1$ as shown in the inset of Fig. 6(a). But the anisotropic case of antiferromagnetic thermal state ($J>0$) is different, the MQC is fragile due to the small gap between $E_0$ and $E_1$ as shown in the inset of Fig. 6(b). Similar conclusions can be obtained when the spin number increases. For the case of ten spins, we further plotted the first ten lowest eigenenergies as the functions of coupling parameter $J$ in Figs. 6(c) and 6(d), which correspond to the isotropic and anisotropic lattices, respectively. For the ferromagnetic thermal state ($J<0$) away from the critical point, the gap between $E_0$ and $E_1$ is very small and the anisotropy has little influence on the two lowest eigenenergies, which gives rise to the fragile MQC at finite temperatures. For the antiferromagnetic case ($J>0$), the anisotropy changes the structure of spectrum as shown in the Figs. 6(c) and 6(d), where the gaps between $E_0$ and $E_1$ have the smaller values in comparison with that of the isotropic case, and then the MQC in the thermal state for the isotropic lattices has the stronger thermal robustness. We further calculate the eigenenergies for the case of 15 spins and obtain the similar conclusion. Moreover, the numerical results show that the MQC $T_{15}$ in the antiferromagnetic regime, its thermal robustness, and the spatial anisotropy of the Ising system still obey the three-way trade-off relations in Figs. 5(d) and 5(e). At the same time, the MQC $T_{15}$ in the ferromagnetic regime is very fragile at finite temperatures.
In Appendix B, we plot the finite temperature effect of MQC $T_N$ for some typical configurations of the triangular lattices with the number of spins being $N=3, 6, 10$, and $15$, respectively.

\begin{figure}
	\epsfig{figure=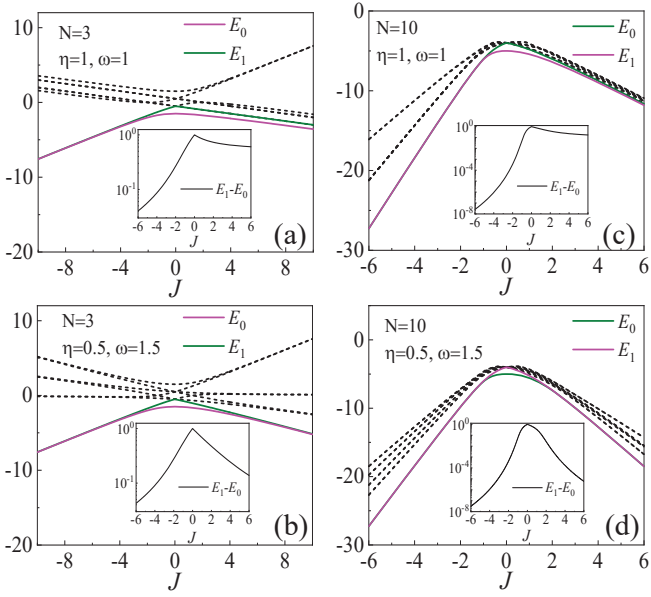,width=0.48\textwidth}
	\caption{(Color online) The energy spectrum for the isotropic and anisotropic Ising systems with the spin number $N=3$ and $N=10$ for which only the first ten lowest eigenenergies are plotted. (a) The isotropic case for $N=3$ with $\omega=\eta=1$. (b) The anisotropic case for $N=3$ with $\eta=0.5, \omega=1.5$. (c) The isotropic case for $N=10$ with $\omega=\eta=1$. (d) The anisotropic case for $N=10$ with $\eta=0.5, \omega=1.5$. The insets in the four panels represent the energy gaps between the ground state and the first excited state.}
\end{figure}

\section{discussion and conclusion}

Since multipartite quantum correlation is an important resource in quantum information processing, it is desirable to study the MQC properties in the triangular Ising system with the tunable interactions. Moreover, in experiments, the tunable spatially anisotropic triangular structure can be realized in various physical platforms, such as trapped ions \cite{kim10nat}, cold atoms \cite{Struck11sci}, artificial-spin-ice heterosystem \cite{wang18natnano}, superconducting systems \cite{niskanen07sci}, and so on. In this paper, we find that unlike the lattices with fixed couplings \cite{rkrao13pra}, the MQC $T_N$ in the antiferromagnetic ground state is sensitive to the spatial anisotropy of the lattices as shown in Fig. 2.

The tuning of spatially anisotropic couplings will provide us an effective strategy for the MQC modulation in the antiferromagnetic ground state. Here, we discuss an experimental possibility for the MQC modulation via the spatially anisotropic coupling in the system of cold atoms. In Ref. \cite{Struck11sci}, the authors utilized the motional degrees of freedom of atoms trapped in a triangular optical lattice to simulate frustrated magnetism. The key technology in their experiment is the independent tuning of the nearest-neighbor couplings $J$ and $J'$ by introducing a fast oscillation of the lattice \cite{eckardt10epl}. The atoms at each site $i$ of the lattice have a local phase $\theta_i$, which can be regarded as a classical vector spin $\textbf{S}_i=[\mbox{cos}(\theta_i), \mbox{sin}(\theta_i)]$. The ground state of this system corresponds to the minimum of the energy \cite{Struck11sci}
\begin{equation}
    E(\{\theta_i\})=-\sum_{\langle i, j\rangle} J_{i j} \mathbf{S}_{i} \cdot \mathbf{S}_{j},
\end{equation}
where the sum extends over all pairs with the nearest-neighbor coupling, and the coefficient $J_{ij}$ has the meaning of coupling strength between the $i$th and $j$th sites. In the experimental system, the sign and magnitude of $J_{ij}$ can be tuned independently via an elliptical shaking of the lattice, which provides a tunable anisotropic parameter $\eta=J/J'$ with the same range of $(0,3)$ as that of our analysis in Fig. 2. Moreover, the recent progress on the tunable couplings \cite{chen14prl,Guo19prl,li20prapp} and the breakthrough in superconducting circuits \cite{Arute19nat,gong21sci,wu21prl,zhang22nat,feng22prapp} can make dozens of qubit be arranged on two-dimensional lattices with tunable nearest-neighbor couplings, which makes it possible to construct an anisotropic triangular structure.

We have looked into the property of the MQC $T_N$ and its thermal robustness in the anisotropic Ising systems at finite temperatures. It is revealed that there exists a three-way trade-off relation among the high MQC, strong thermal robustness and the anisotropic strength in the triangular Ising system with antiferromagnetic interactions as shown in Fig. 5. In particular, one can obtain the relatively high MQC and strong thermal robustness at the same time via tuning the spatially anisotropic interactions in the antiferromagnetic Ising lattices. The trade-off relation originates from the structure of the first two lowest eigenenergies as shown in Fig. 6, which has close relation with the anisotropy and gives rise to different kinds of thermal states. The qualitative analysis still holds for the lattices of more spins. Experimentally, the tunable triangular lattices can be utilized as a basic building unit for constructing complex multipartite quantum networks \cite{wehner18sci,zhong21nat} and served as a potential platform for detecting quantum nonlocality in many-body systems \cite{luo18prl,renou19prl,tejada21prl}.

In conclusion, we have investigated the multipartite quantum correlations (MQCs), spatially anisotropic coupling, and finite temperature effects in triangular Ising systems with tunable interactions. In the ground states of the Ising lattices, the influence of anisotropy on MQC varies, allowing us to modulate MQC in the antiferromagnetic ground state using anisotropic strength. This modulation is feasible with current experimental technologies. Moreover, we have uncovered a three-way trade-off relationship between MQC, thermal robustness, and anisotropic coupling in the tunable Ising system with antiferromagnetic couplings, while MQC in the ferromagnetic case is highly susceptible to finite temperatures. Our findings offer valuable insights into ground state properties and MQC modulation in quantum many-body systems.

\begin{acknowledgments}
This work was supported by the NSF-China (Grant Nos. 11575051, 11904078, 12074376 and 12105074), Hebei NSF (Grant Nos. A2021205020, A2019205263 and A2019205266), and the CRF of Hong Kong (C6009-20G). JR was also funded by the project of China Postdoctoral Science Foundation (Grant No. 2020M670683).
\end{acknowledgments}

\appendix

\section{finite temperature effect of two-spin negativities in the tunable lattices of ten spins}

\begin{figure}
	\epsfig{figure=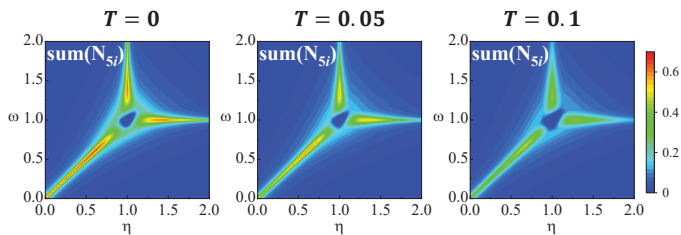,width=0.5\textwidth}
	\caption{(Color online) The two-spin negativities $\sum_i N_{5i}$ as a function of parameters $\omega$ and $\eta$ and its thermal robustness in the spatially anisotropic triangular lattices, where the temperature is chosen to be $T=0, 0.05$ and $0.1$, respectively.}
\end{figure}

According to the analysis in Sec. III of main text, we know that the MQC in the isotropic triangular Ising structure has the strongest thermal robustness and there exists a three-way trade-off relation among the MQC, its thermal robustness and the spatial anisotropy. However, the behavior of two-spin negativities is still not clear. Here, we consider the tunable Ising lattices of ten spins and analyze the relation among two-spin negativities $\sum_i N_{5i}$, its thermal robustness and the spatial anisotropy. In Fig. 7, we plot the negativities $\sum_i N_{5i}$ as a function of anisotropic parameters $\omega$ and $\eta$ at the temperature $T=0, 0.05$, and $0.1$, respectively.

In comparison with the MQC in Fig. 5(c), the behavior of negativities $\sum_i N_{5i}$ is quite different and, in particular, $\sum_i N_{5i}$ has almost zero values for the configurations close to isotropic lattices. At zero temperature, the higher negativities exist in the configuration close to weak spatial anisotropic lattices (\emph{i.e.}, the lattices with two smaller equal couplings) as shown in Fig. 7(a). Along with the increase of temperature, the two-spin correlations have the strong thermal robustness for the configurations of weak spatial anisotropy as shown in Figs. 7(b) and 7(c). Moreover, we further study the two-spin negativities $\sum_i N_{5i}$ in the tunable Ising lattice of $15$ spins and can obtain the similar results.

\section{finite temperature effect of the MQC for some typical configurations}

\begin{figure}
	\epsfig{figure=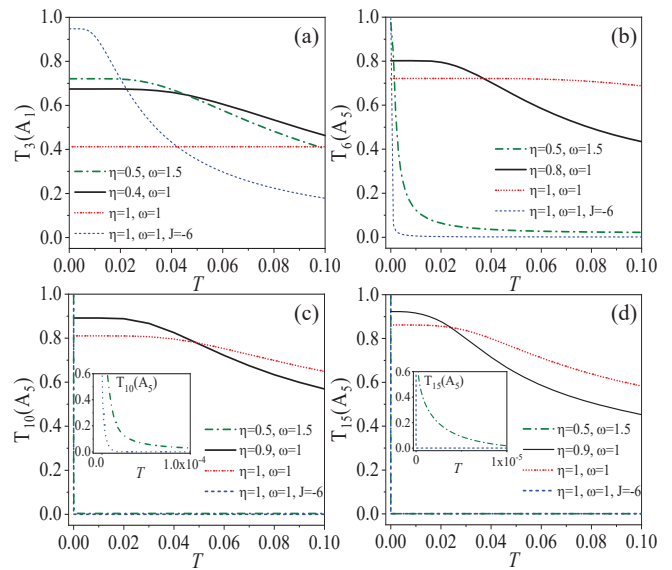,width=0.48\textwidth}
	\caption{(Color online) Multipartite quantum correlation $T_N$ as a function of temperature $T$ with different parameters of the tunable interactions in ferromagnetic regime (blue dashed line) and antiferromagnetic (green dash-dotted line, black solid line, and red dotted line) regimes. Panels (a)-(d) correspond to the case of spin numbers $N=3, 6, 10$, and $15$, respectively. The insets in panels (c) and (d) highlight the MQC in the ferromagnetic regime and that in the antiferromagnetic regime with the strong spatial anisotropy.}
\end{figure}

In order to further illustrate the relation between the MQC $T_N$ and the anisotropic configuration of the lattices at finite temperatures, we plot $T_N$ as a function of the temperature $T$ for some typical configurations with the spin numbers $N=3, 6, 10$, and $15$, respectively. As shown in Fig. 8, for the ferromagnetic regime  ($J=-6$), we consider the isotropic lattices with the parameters $\eta=\omega=1$ (blue dashed line), and for the antiferromagnetic regime ($J=6$), we choose three typical configurations, \emph{i.e.}, the strong spatial anisotropy ($\eta=0.5$, $\omega=1.5$, green dash-dotted line), the weak spatial anisotropy ($\eta=0.9$, $\omega=1$, green black solid line), and the isotropy case ($\eta=\omega=1$, red dotted line).

In Fig. 8(a), the three-spin case is plotted, where the MQC $T_3$ in the ferromagnetic regime (blue dashed line) decays fastest and the one with isotropic structure (red dotted line) in the antiferromagnetic regime has the strongest thermal robustness. In addition, the strong anisotropic case (green dash-dotted line) has the higher initial $T_3$ than that of the weak anisotropic one (black solid line) but decays much faster. Along with the number of spins increases, we find that the decay of $T_N$ in the ferromagnetic regime (blue dashed line) speeds up as shown in Figs. 8(b)-(d), and the case for the strong anisotropic structure (green dash-dotted line) is similar, for which the reason can be explained by the analysis on structure of energy spectrum in Fig. 6 of main text. Moreover, the MQC $T_N$ for the weak anisotropy (black solid line) has higher initial correlation than that of the isotropic case (red dotted line) but decays a little faster along with temperature. The numerical results for the finite temperature effects of $T_N$ with $N=3,6,10,15$ coincide with our previous theoretical analysis in Sec. III of main text.

\end{document}